\documentclass[runningheads]{llncs}
\usepackage{graphicx}
\usepackage{xcolor}
\usepackage{url}
\usepackage{float}
\usepackage{soul}
\usepackage{todonotes} 

\newcommand{\ctext}[3][RGB]{%
  \begingroup
  \definecolor{hlcolor}{#1}{#2}\sethlcolor{hlcolor}%
  \hl{#3}%
  \endgroup
}
\begin{document}

\newcommand{\jenote}[1]{\ctext[RGB]{0,209,82}{~#1~}}
\title{Towards External Calls for Blockchain\\and Distributed Ledger Technology}
%
%
\author{Joshua Ellul \and Gordon J. Pace}
\institute{Centre for DLT\\
and Department of Computer Science\\
University of Malta\\
\email{joshua.ellul@um.edu.mt}\\
\email{gordon.pace@um.edu.mt}\\
}

\authorrunning{Joshua Ellul and Gordon J. Pace}
%

%
\maketitle              
\begin{abstract}
It is widely accepted that blockchain systems cannot execute calls to external systems or services due to each node having to reach a deterministic state. However, in this paper we show that this belief is preconceived by demonstrating a method that enables blockchain and distributed ledger technologies to perform calls to external systems initiated from the blockchain/DLT itself.
\end{abstract}

\section{Introduction}
Bitcoin \cite{nakamoto2008peer}, a `peer-to-peer electronic cash system', brought about a wave of interest to not only enable peer-to-peer payments but also to decentralise control of assets. A few years later, the publication of the Ethereum white paper~\cite{wood2014ethereum} introduced a way to decentralise computation through smart contracts, an unfortunately chosen term which led to much confusion (and undue comparison) with legal contracts, which enabled for decentralisation of computational digital processes. Ethereum further extended the wave of interest beyond that of decentralised control to one of decentralised logic. 

The last few years have seen many attempts to decentralise various applications, and whilst the blockchain sector is healthy \cite{ellul2021blockchaindead}, a major problem remains that of mass-adoption \cite{prewett2020blockchain,stratopoulos2020blockchain}. The majority of potential users are not tech-savvy and neither educated in regards to the benefits of decentralised systems, and until such users become educated and build trust in such systems and until sufficient tooling is provided to allow for the non-tech savvy to use such systems, then mass-adoption will be difficult to achieve.

Many have argued that decentralisation is a cure to many woes arising from issues of trust. By removing centralised points-of-trust, one can build solutions which empower all participants. Or so some have argued. Blockchain (and other distributed ledger technologies) allow for such decentralisation and provide an opportunity to build fairer asset management, or more generally services for the users. There may be truth in such statements, but one thing that this view fails to take into account is that the real world lies outside the blockchain, and although data and algorithms residing on the blockchain can be decentralised, any reference to the real world must necessarily break through the event-horizon of the blockchain and interact with the outside world much of which is centralised out of physical or regulatory necessity. For instance, if one needs to access the temperature at a particular location at a particular time or if one wants to actuate a switch to unlock a door, one must interact with the real world and trust that the correct information has been provided or that the right instructions have been carried out.

Blockchain systems have traditionally addressed these issues through the use of oracles --- channels providing information from the outside world to the blockchain. However, the nature of public blockchains allows only for a one way flow of information --- from external entities into the blockchain --- and any attempt to do this in the opposite direction (i.e. invoke an external entity from within the blockchain) causes problems due to the nature of consensus on such systems\footnote{We discuss this in more detail in Section~\ref{ssec:technicalchallenges}.}. The only alternative solutions available require trusted entities which perform such invocations, which simply delegates the problem one step away.


In this paper, we present a solution to enable direct and seamless communication between blockchain systems with external systems back and forth, which we believe will enable for new types of decentralised applications to emerge (and make more strides towards achieving mass-adoption).

The paper is organised as follows. In Section~\ref{sec:motivation} we motivate further the need for direct calls emanating from public blockchains and present the technical challenges which have stood in the way of building solutions to this problem till now. We then explain the solution we are proposing in the remaining sections. Section~\ref{sec:ver} explain the encoding of verifiable external calls, Section~\ref{sec:tx} presents the extended structure of transactions to handle them and its impact on blocks in Section~\ref{sec:blk}. Finally, Section~\ref{sec:net} brings these together at the level of the network.


\section{Motivation and Challenges}
\label{sec:motivation}
The current status quo of public blockchains has certain limitations which would be addressed by enabling direct external calls, but doing so is fraught with technical challenges. We discuss these and explain how our contribution fits in this section.

\subsection{On Why External Calls}

The work presented herein is motivated by the following two objectives: (i) to enable for more efficient dApp processes; and (ii) to allow for applications that require direct blockchain to external systems to be realised (some of which can help work towards mass-adoption). Indeed, we believe this approach will enable for much more than just these two objectives which we now discuss further below.

\textbf{More efficient dApp processes}: It is becoming evident that trusted external oracle input is essential for many types of dApps being proposed --- e.g. within the DeFi (Decentralised Finance) space. DeFi is just one use-case that highlights the importance that external trusted party data plays in the decentralised world and, whilst, the state-of-the-art does provide a solution to providing this required input, it is worth investigating whether direct blockchain to oracle interaction can be achieved in aim of providing more efficient processes. 

\textbf{New applications that require external calls}: Whilst there is a long standing debate in regards to whether regulation of the crypto-space defies the whole point of such self-regulating systems, and without getting into the debate, we do acknowledge that regulation and legal certainty may also help bring about mass-adoption since many non-tech savvy users may trust jurisdictional authorities over code. However, regulatory and legal requirements of such technology raises problems, as requirements often require mandatory interaction with regulatory entities or other parties (e.g. that accounts must be filed at the end of the year). This creates problems for such decentralised (potentially autonomous) organisations. Whilst, initial thoughts towards creating legal certainty for such types of organisations have been presented \cite{ganado2020mapping}, such mandated direct interaction with the external world from blockchain systems, smart contracts, decentralised autonomous organisations (DAO) or any other decentralised logic was not possible prior to the work being proposed herein. Indeed, this is just one such use-case in regards to types of applications (regulated ones that require direct communication with external parties) which this work can enable, and we believe that many other applications which would require a blockchain/DLT to communicate directly with the external world will be proposed based on this work.



What is needed is a decentralised system that supports both decentralised peer-to-peer transactions and logic but also the ability to interact with external trusted parties directly. Decentralised application frameworks and smart contracts enable a way to achieve passive interaction with external parties, however suffer from obligating the external points-of-trust to integrate and actively interact with the respective dApps at their cost. This is not always feasible since such points-of-trust (such as certificate providers, universities, authorities, regulators, distributors, and any other centralised service provider) often follow rigid processes and would likely be unable to integrate directly with a dApps's technical integration requirements. Even more so when such points-of-trust may be responsible for serving (or monitoring) many operators --- which would potentially mean that they would need to integrate with many smart contracts/dApps/blockchain platforms.

\subsection{The Challenges}
\label{ssec:technicalchallenges}

Let us start by looking at technical challenges that are in the way of achieving feasible and efficient oracle input. Blockchain and Distributed Ledger Technology (DLT) systems require that the decentralised logic encoded within them reaches a deterministic state. It is said that every node must execute the exact same logic in order to achieve consensus.   

For this reason, it is the general consensus in the community that Blockchain and DLT systems cannot make calls to external systems/oracles \cite{ellis2017chainlink,caitruth,grootemanproviding,greenspan2016many,molinabenefits,ma2019reliable,da2019trustable,dinh2019blueprint,muhlberger2020foundational,adler2018astraea,kamiya2019shintaku,guarnizo2019pdfs,woo2020distributed,van2018publicly}. External data is fed in from oracles who actively issue transactions to the DLT --- which means that the oracle must pay for providing this information. Many oracles that could provide useful information may not be incentivised to do so --- why should a external party provide information at a cost if there is nothing in it for them? Solutions to this problem have been proposed which require external systems and networks to feed such oracle input into smart contracts \cite{al2020trustworthy} --- again, however, requiring that such costs are borne (typically by the end-users or dApp operator). Also, feeding in oracle input imposes delays, some which may be due to the oracles' own delays in submitting required transactions, and others due to the fact that different steps within an interaction will take place in different blocks.

For small private DLT systems, a solution to initiate externals calls to trusted parties had been proposed\footnote{\label{fn:hyper}https://hyperledger.github.io/composer/v0.19/integrating/call-out} which requires that each node makes the same call to the external system and receives the exact same response. While this may work for smaller sized networks (though in an inefficient manner), the solution does not scale up as networks increase in size. More so, this solution limits the types of calls that can be made to services that provide the same response irrespective of when the call is made --- i.e. stateless and deterministic. The reason is that all nodes must receive the same response, and also that since blocks may need verifying at a later point in time, the same response would need to be received and verified in future as well. Therefore, such a solution could limit the extent to which the chain could be verified if blocks ever need to be verified in future and the external service provider is either no longer available or returns different responses to the original response received. 
More so, it would be impossible in future to determine whether the participating verifying nodes actually recorded the correct response, or whether they had all agreed on an incorrect response. Whilst, approaches for non-interactive deterministic communication have been proposed\footnote{https://ethresear.ch/t/on-chain-non-interactive-data-availability-proofs/5715} \cite{adler2019building}, to the authors' best knowledge the work presented herein is the first proposed work on active communication which maintains deterministic computation across the network.

So far dApps have required that any parties directly interact with blockchain networks (or make use of proxy services \cite{zhang2016town}). This is not ideal as it limits parties: (i) to those that are willing to interact and integrate with the specific smart contract/network and bear any costs required to initiate transactions; or (ii) by requiring dApps to make use of external services which become trusted parties themselves. Chainlink \cite{ellis2017chainlink} is such a trusted party which makes use of decentralised nodes to lower trust issues in single node operators as well as provides a bridge between various data sources and blockchain systems --- yet it requires fees to be paid for providing the service\footnote{https://medium.com/@chainlinkgod/scaling-chainlink-in-2020-371ce24b4f31}. 

\begin{figure}[H]
\centering
\includegraphics[scale=0.85]{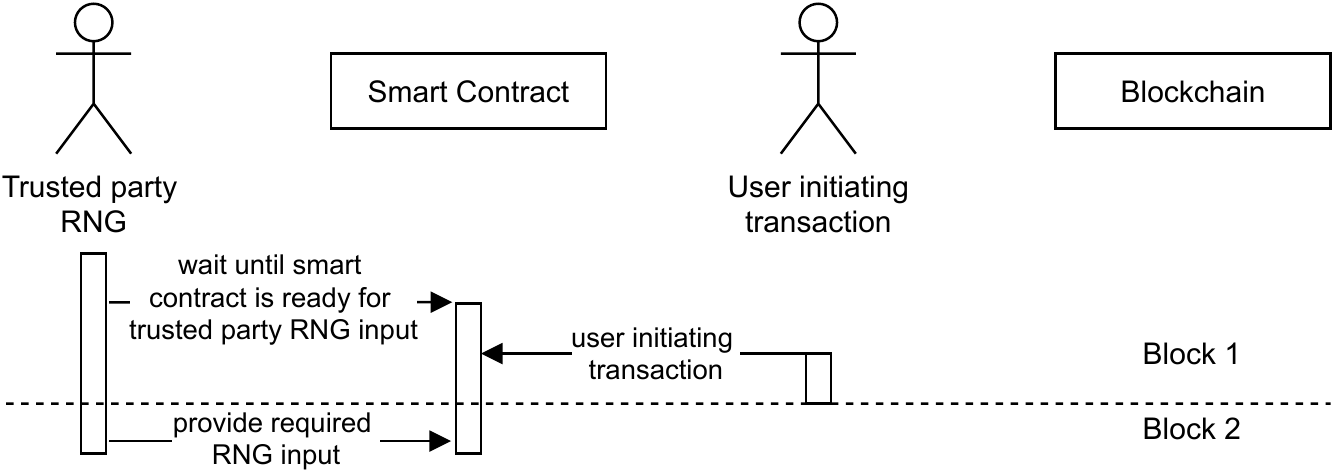}
\def\svgwidth{\columnwidth}
\caption{Trusted parties may often require to wait for an action in a first block before it can transmit its input into a subsequent block.}
\label{fig:rng}
\end{figure}

Any required input from trusted parties results in separate transactions belonging to separate blocks which could result in potentially large delays for processes making use of oracles. Such delays can be large when trusted parties are actively monitoring the ledger, let alone when trusted external parties are not listening for such changes in real-time. As an example consider a case where a trusted party random number generator (RNG) is used to input a random number which is only revealed after a user initiates the smart contract process --- perhaps the user is placing a bet in a decentralised betting dApp. The random number should not be revealed before the user places the bet since the user would be able to know the outcome before betting. This first transaction by the user to place a bet would take place in Block 1. Thereafter, at best, the input from the RNG trusted party can take place in Block 2 (though the input transaction could also be initiated and/or be accepted in later blocks). The following sequence diagram in Figure~\ref{fig:rng} depicts this.

\subsection{Contribution Overview}
To recap, trusted oracles are key for types of decentralised services that interact with the real-world. The general consensus is that it is impossible or infeasible to allow blockchain systems to make direct calls to external parties due to: (i) the requirement for computation to reach a deterministic state; and (ii) potential overloading of points-of-trust with requests. Also, whenever a node requires to verify some computation which involves a point-of-trust, it would have to send another request, possibly getting a different result than what the original caller got --- which would also inundate the centralised node with requests (potentially resulting in a denial-of-service attack). 

\begin{figure}[t]
\centering
\includegraphics[scale=0.90]{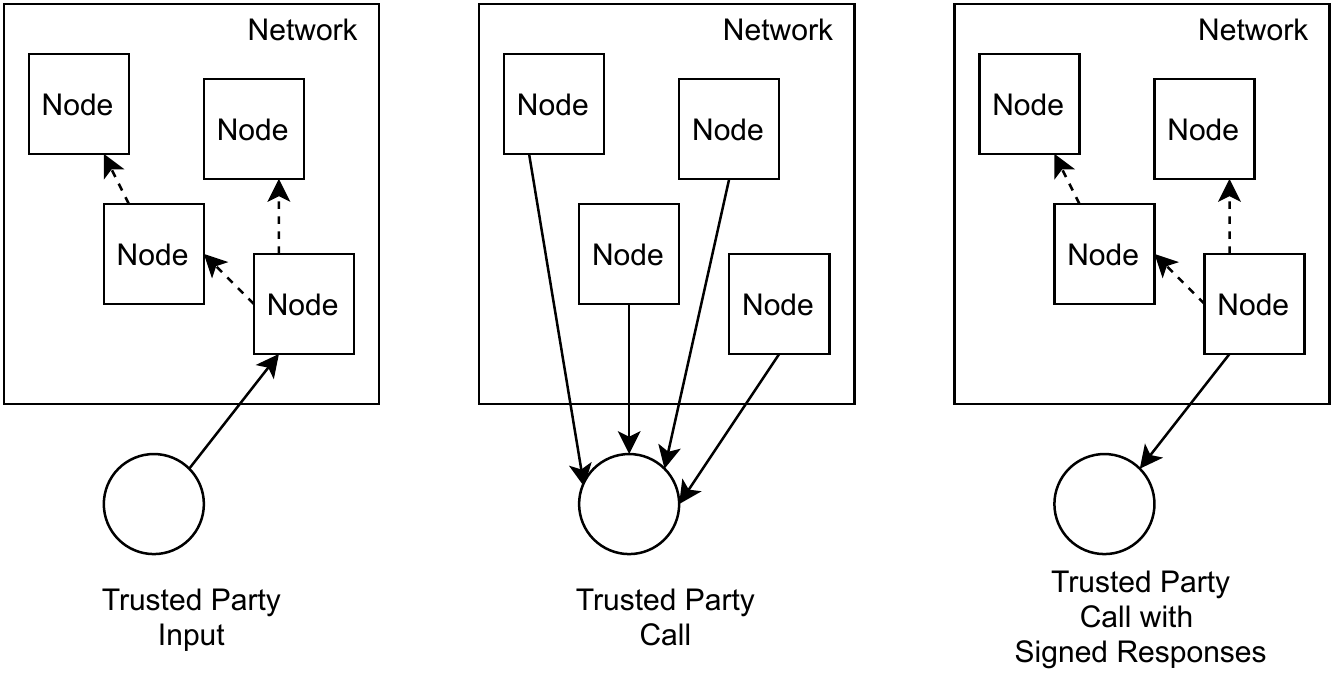}
\caption{Left: traditional trusted party input; middle: active calls requiring each node to undertake the external call that must return the same input; right: external calls enabled with verifiable signed responses.}
\label{fig:signedresp}
\end{figure}

However, we believe that the general consensus on this matter is not well founded and preconceived. Perhaps based upon the often cited deterministic nature of computation that is required \cite{sankar2017survey} --- yet whilst this statement is true, it is important to highlight that it is the state which computation reaches that must be deterministic, and the computation performed can reach such a deterministic state in different ways. In this paper we present a mechanism that allows for such systems to interact with external parties directly in a feasible manner. Figure~\ref{fig:signedresp} provides an overview of an oracle input transaction/call flow for: (i) traditional oracle input (on the left); (ii) (inefficient) external calls requiring responses to always be the same --- which does not scale up (in the middle); and (iii) the solution proposed herein which makes use of verifiable external calls (on the right). 

\section{Verifiable External Calls}
\label{sec:ver}
The solution proposed herein is to make use of verifiable external calls --- i.e. a request (call) made to an external system that returns back a signed response which: (i) can be verified to truly be a response from the external party in question; (ii) which does not require any further communication (with the external party or other). This can be achieved in the same way how we provide such assurances in traditional applications and how trusted oracle input is verified, by checking whether the response was indeed digitally signed \cite{merkle1989certified} by the external party. Knowledge of the trusted party's public key is required to be known (in the same way that oracle input requires knowledge of the trusted party's address) or can be retrieved from a trusted entity.

To allow for processes to make direct use of external services (in a feasible and efficient manner), which do not require explicit integration from the external parties themselves (with the specific platform), we propose to make use of verifiable external calls which provide a guarantee with respect to the veracity of the origin of the response both at the time of processing as well as for any point in future for which such verification may be required. 

A verifiable external call is defined as the following tuple --- a request, a public key and a signed response structured as follows:
\[ \langle request, \;public\_key,\; signed\_response \rangle \]

The $request$ would likely be represented as a Uniform Resource Identifier (URI) and should point to the external system/service endpoint which is to be called (though this is an implementation design decision), and may also comprise of other input data. The $public\_key$ may be hard-coded into the application logic (e.g. into the smart contract), or it could even be retrieved by a trusted certificate provider. In either case it would need to be recorded by the time when the external call is executed --- it will be used to verify the response originated from the respective external party. The $signed\_response$ is the response that has been signed using the external party's private key (which is associated with $public\_key$.

 Indeed, this does require that the trusted data sources provide an end-point that responds back with a signed response which would likely require changes to existing data sources to implement signed responses --- however, recent proposals indicate that such a standard may eventually be adopted\footnote{\url{https://wicg.github.io/webpackage/draft-yasskin-http-origin-signed-responses.html}}, which if adopted would enable for this approach to integrate with all data sources (that are keeping up with standards). 

Furthermore, to avoid old signed responses from being repeated, an incremental number, date/time or challenge-response could be made use of which would ensure old responses cannot be repeated --- however this is left as an implementation detail. Whilst, the challenge data sent to the external party will be part of the $request$, the verifiable external call's definition may be extended to include the challenge-response. For example, the request can be augmented by a request number to a fresh nonce $\nu$, which is expected to be included unchanged in the response:
\small\[ \langle request \oplus \{\textit{request\_nonce}\mapsto\nu\}, \;public\_key,\; signed\_response\oplus \{\textit{response\_nonce}\mapsto\nu\} \rangle \]\normalsize

\section{Transactions}
\label{sec:tx}
\begin{figure}[t]
\centering
\includegraphics[scale=0.9]{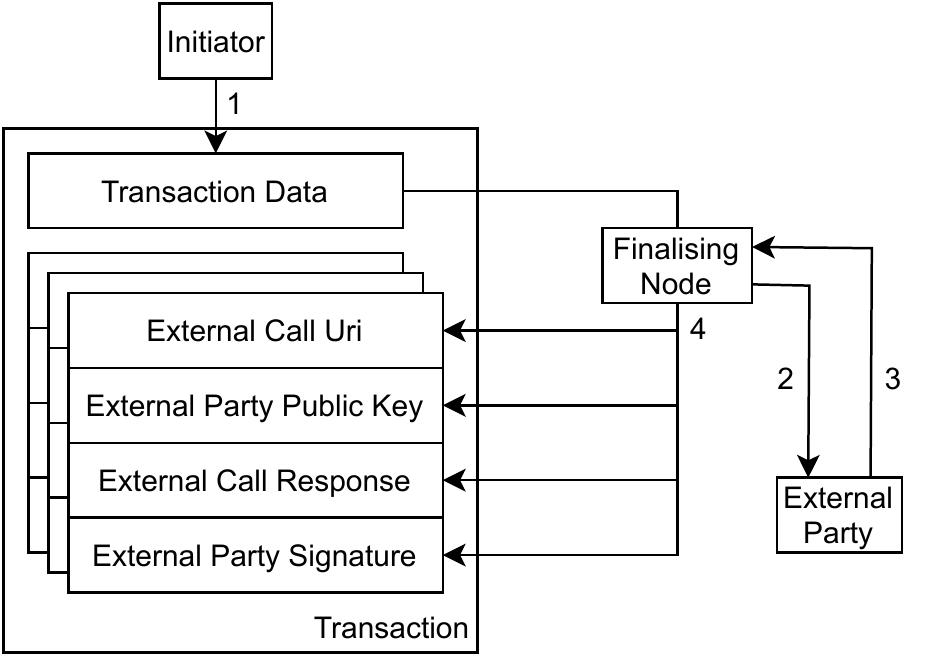}
\caption{Transaction finalisation process.}
\label{fig:trans}
\end{figure}

When a transaction is initiated (be it by a user, another system, or the system itself if such a DLT allows this) and accepted for execution, the node which is processing the transaction will establish all external calls which need to be performed, execute them and record the responses received back from the trusted external parties along with associated digital signatures. Indeed, at this point the finalising node must ensure that the response is from the trusted party by verifying the response and signature against the trusted party's public key. Furthermore, if a unique number, date/time, or challenge-response mechanism was used to ensure old data is not repeated, then this would also be validated at this point. A depiction of how a transaction is initiated and attributed with the various data associated with external calls is depicted in Figure~\ref{fig:trans}.

If any responses are not verified, then the transaction may be deemed to have failed, or depending upon reparation or other logic the transaction may still be valid (and able to process the unverified response). This is a design decision that each platform would need to consider. The same goes for external calls for which no response is received. 

To reiterate, to ensure that a finalising node does not repeat old responses from external parties, the response and signature could be accompanied with the date and time the response was generated and/or a unique response identifier associated with the response (and potentially request as well). One challenge is to ensure that participating nodes indeed execute such external calls rather than simply record failure, which we will delve into in a future paper (since it merits its own paper). However, verifiable external calls could also be undertaken by the transaction initiator (i.e. the party submitting the transaction provides this information as part of the transaction submission process) --- however indeed this depends upon the architectural design of the blockchain, smart contract and wallet/dApp software submitting the transaction. By performing the verifiable external call at transaction submission time (on the initiator), the aforementioned problem pertaining to nodes potentially reporting back failed external calls would be eliminated. 

\section{Construction of Blocks}
\label{sec:blk}
Block construction can proceed in the conventional manner, adding all transactions including external invocations and responses as part of the block. However, depending upon the type of system, it may be the case that a particular trusted party is used extensively for multiple transactions within the same block and that same response would be also be received for the various other transactions in the block. For such systems one can optimise-out redundant external calls by implementing a response cache which keeps track of responses previously received within the same block and uses these values if required again. Such an approach would necessarily assume that the external calls are side effect free. This is obviously application dependent and therefore well-defined semantics should be in place in regards to whether a cached response may be used, or whether a new response must always be retrieved for different transactions --- and should be clear to the various system's stakeholders and users.

In fact one can go one step further in the use of this caching mechanism by allowing the use of cached responses in older blocks to be referenced and reused in new transactions. Needless to say,  consideration would need to be made in regards to whether the particular system would allow for historical data lookup. 

\begin{figure}[t]
\centering
\includegraphics[scale=0.85]{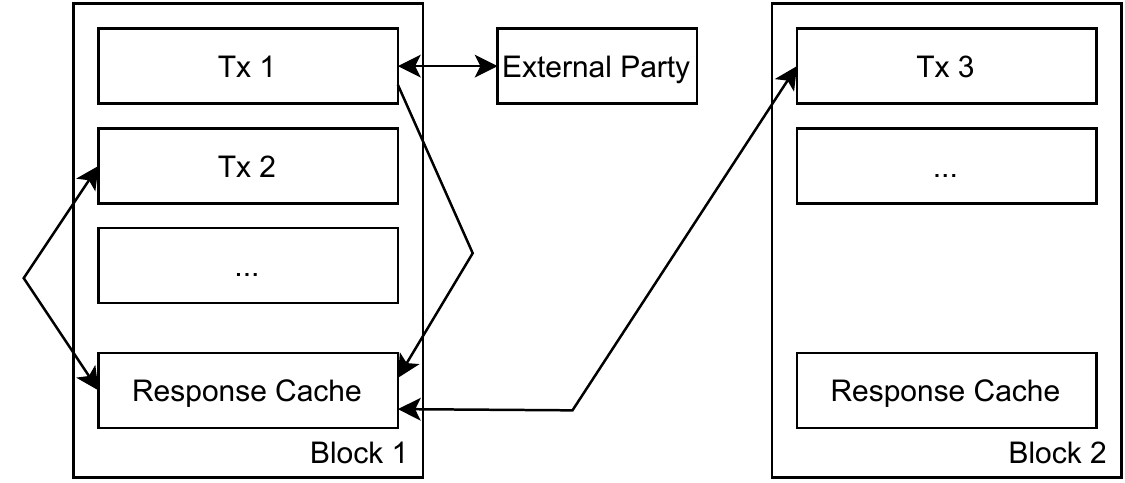}
\caption{Depiction of response cache usage in same and different blocks.}
\label{fig:blocks}
\end{figure}

The diagram in Figure~\ref{fig:blocks} illustrates the use of external call caches. Two transactions in \textit{Block-1} (\textit{Tx-1} and \textit{Tx-2}) make use of the same external call. If \textit{Tx-1} is the first transaction of the two to be executed, the finalising node executes the external call and records the signed response in the response cache, referring to it in \textit{Tx-1}. When \textit{Tx-2} is executed, it finds that the call has already been recorded in the response cache (since we are assuming that it makes use of the same call as \textit{Tx-1}) and uses the cached value instead of invoking the external call again. Similarly, in \textit{Block-2}, the inclusion of transaction \textit{Tx-3} which also matches the call in the previous block can also make use of the same cached signed response stored in that cache.


Limitations on the number of external calls that can be issued per transaction\footnote{A transaction invoking a call to a smart contract can result in multiple external calls arising from that single transaction.} and potentially per block may need to be established explicitly or implicitly (for example through a `gas'-like feature) to avoid delaying block generation. One potential design feature may be to implement transactions which involve various external calls to span across blocks, allowing for some logic and external calls to be issued in one block to be followed by the continuation of the execution of the logic and possibly more external calls in a later block. It is worth noting that if one were to adopt verifiable external calls executed by the transaction initiator (see previous section), no such considerations are necessary.

\section{Network and Consensus}
\label{sec:net}
At a network level, one may na\"{i}vely keep existing approaches allowing the node finalising the block to broadcast it once complete. However, this would result in a multitude of nodes attempting to include transaction which include external calls to perform the invocations, thus flooding the service provider with calls. Instead, we propose a two-staged approach in which a node would first bid for the right to finalise a node\footnote{We refer to bidding and earning the right to finalise a block in order to remain agnostic to the underlying consensus algorithm.} without having performed the external calls and only then, after earning this right, would it perform the calls and submit the concluded block to the network. For instance, in a proof-of-work setting, a bid for a block would include all the block information \emph{except for the transaction response cache}. Consensus would then proceed as usual, with other nodes accept this partial block if it conforms to the usual proof-of-work requirements. Once such a partial node is accepted (i.e. added to the chain), the node which won can perform the external calls and submit the signed results to be added to the chain as the end of that block. Other nodes would accept the submitted results only if they are duly signed by the invoked service provider. Consideration would have to be taken to take into account not receiving the end of the block from the winning node.

\begin{figure}[t]
\centering
\includegraphics[scale=0.95]{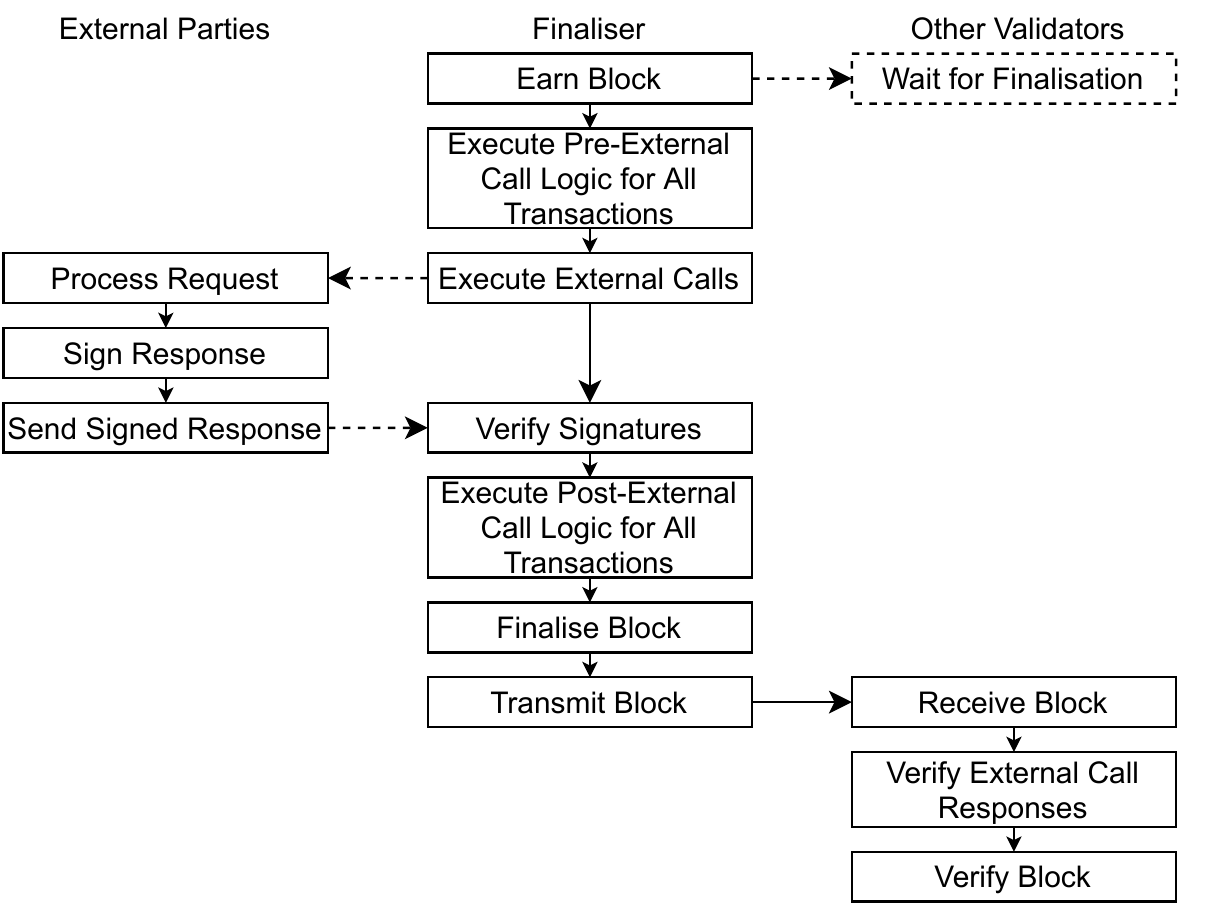}
\caption{Network-level block finalisation overview.}
\label{fig:network}
\end{figure}

In general, this would work as follows: (i) When a node has `earned' the right to finalise a block (irrespective of how this is established) the finalising node may send out the block solution (if required) to let other nodes know that a solution has been found so that they can decide whether to wait for the block or continue looking for a solution; (ii) in such cases it would be up to each node implementation to decide how long to potentially wait for a block until continuing to look for a new one; (iii) for algorithms where the finalising node (miner/validator) is determined implicitly this would not be required. An overview of the process is depicted in Figure~\ref{fig:network}

When executing transactions, finalising nodes would thus execute up to the point where external calls are required to be resolved, in which case the finalising node should make the requests, receive the responses and verify that they have been signed by the respective trusted party. Thereafter, remaining logic pertaining to the transactions can be executed. The block can then be finalised and sent to other nodes, who will in turn verify that the responses received are from the trusted parties. 
Once again such considerations are not required for initiator-based transactions.

\section{Conclusions}
It is a widely accepted belief that blockchain systems cannot execute calls to external systems due to the requirement for computation to reach a deterministic state. It is often said that blockchain-based computation needs to be deterministic \cite{sankar2017survey}, however it is important to highlight that requiring the computation to reach a deterministic state does not mean that different computation cannot take place to reach that state. Contrary to the general consensus, in this paper, we have demonstrated a method for Blockchain and DLT systems that allows for direct external calls to be initiated from the Blockchain/DLT itself (or even from a transaction initiator's software if suitable). 

A prototype demonstrating verifiable external calls has been implemented and available from \url{https://github.com/joshuaellul/excalls}, and further information on the project is available from: \url{https://joshuaellul.github.io/excalls/}  (which will be updated with progress).

\bibliographystyle{splncs04}
\bibliography{refs}

\end{document}